# Nanoparticle-lipid interaction:
# Job scattering plots to differentiate vesicle aggregation from supported lipid bilayer formation

F. Mousseau[1], E.K. Oikonomou[1], V. Baldim[1], S. Mornet[2] and J.-F. Berret[*1]

[1]*Matière et Systèmes Complexes, UMR 7057 CNRS Université Denis Diderot Paris-VII, Bâtiment Condorcet, 10 rue Alice Domon et Léonie Duquet, 75205 Paris, France.*

[2]*Institut de Chimie de la Matière Condensée de Bordeaux, UPR CNRS 9048, Université Bordeaux 1, 87 Avenue du Docteur A. Schweitzer, Pessac cedex F-33608, France*

**Abstract:**
The impact of nanomaterials on lung fluids or on the plasma membrane of living cells has prompted researchers to examine the interactions between nanoparticles and lipid vesicles. Recent studies have shown that nanoparticle-lipid interaction leads to a broad range of structures including supported lipid bilayers (SLB), particles adsorbed at the surface or internalized inside vesicles, and mixed aggregates. Today, there is a need to have simple protocols that can readily assess the nature of structures obtained from particles and vesicles. Here we apply the method of continuous variation for measuring Job scattering plots and provide analytical expressions for the scattering intensity in various scenarios. The result that emerges from the comparison between modeling and experimental measurements is that electrostatics plays a key role in the association, but it is not sufficient to induce the formation of supported lipid bilayers.

**Keywords**: Nanoparticles – Bio-nano interfaces – Electrostatic interactions – Supported Lipid Bilayers

Corresponding authors: jean-francois.berret@univ-paris-diderot.fr
To be submitted to Colloids and Interfaces
Thursday, September 20, 18

# I – Introduction

The emission of fine and ultrafine particulate matter in the environment is responsible for the increase of mortality and morbidity from cardiorespiratory diseases worldwide [1-2]. In the context of environmental pollution, engineered nanoparticles which sizes are less than 100 nm have attracted much attention and been identified as potentially harmful. When inhaled, these particles are able to reach the respiratory zone in the lungs and enter in contact with the alcinar region composed of hundreds of millions of alveoli [3-4]. Several scenarios of nanoparticles passing from the alveolar spaces towards the blood circulation have been examined recently and in some case studies the crossing of the air-blood barrier has been demonstrated [4]. It is found that in the alveolar spaces, the nanoparticles first come into contact with the pulmonary surfactant, a fluid composed of lipids (90%) and proteins (10%) which provides important functions in the lung physiology [5-6]. This scenario prompted researchers to actively study the interactions between nanoparticles and lipid vesicles, typically with vesicular structures in the size range 100 nm to 1 μm [7-10].

Another example where particles interact directly with biological membranes is the process of endocytosis [11]. Endocytosis is the biological process by which nano-objects of different nature and sizes, including pathogens, bacteria, virus, nanoparticles etc… are internalized inside living cells. For particles larger than 1 μm, the process is referred to micropinocytosis, whereas for 100 nm nanoparticles the passage through the membrane can be passive or active, this later being mediated by caveolin or clathrin proteins [12]. When nanoparticles are close to a cell membrane, the interactions generate forces of different origins (e.g. van der Waals, electrostatic), leading to the membrane wrapping around the particles and cellular uptake [13-15].

To evaluate the interplay of nanomaterials with biological membranes, recent experimental, theoretical and simulation studies have focused on the interaction of nanoparticles with closed membranes in the form of vesicles [9,16-23]. Most experiments reported until recently were performed using synthetic lipids of the phosphatidylcholine class, such as 1,2-dipalmitoylphosphatidylcholine (DPPC), 1,2-dimyristoyl-*sn*-glycero-3-phosphocholine (DMPC) or 1,2-dioleoyl-*sn*-glycero-3-phosphocholine (DOPC). From their structures, the phosphatidylcholine vesicles are similar to those found in pulmonary surfactants [5,17,24-26]. Strategies based on the use of more biological models and substitutes have been also proposed [17,27-29]. Depending on the particle size, charge and hydrophobicity, several mechanisms have been suggested, leading to a wide variety of hybrid structures. Fig. 1 displays a library of nanoparticle-membrane structures observed using cryogenic transmission electron microscopy (cryo-TEM). They include nanoparticles coated with a single bilayer (called nano-SLB in the following, Fig. 1a) [17,28], particles embedded with the lipid membrane or adsorbed at the surface (Fig. 1b and 1c respectively) [30-31], particles internalized inside the lipid compartment (Fig. 1d) [32] and mixed nanoparticle-vesicle aggregates (Fig. 1e) [18]. In the case of particle internalization, the fluid membrane invaginates and envelops one or several particles like in cellular endocytosis [11]. Despite many efforts, the mechanisms of particles interacting with the synthetic or biological membranes are not fully understood.

A broad range of experiments was used to study particle-membrane interaction. These experiments include, among others light scattering [17,33], leakage assays [18], quartz crystal microbalance [34], electron and fluorescence microscopy [17,24-25,30-32]. Cryogenic transmission electron microscopy is probably one of the best methods to visualize the nanoparticle-membrane structures, as illustrated in Fig. 1 [35]. It has the required resolution (~ 1 nm) and electronic contrast to identify both nanoparticles and lipid membranes. Cryo-TEM images however lack of statistics, as only a few objects are usually displayed. In this context, there is need to develop simple protocols that can rapidly assess the nature of structures obtained from particles and vesicles. Here we provide examples of light scattering based analytical models that are able to discriminate among the different association scenarios illustrated in Fig. 1. The approach is developed for static light scattering but could be extended as well to small-angle neutron and X-ray scattering or UV-visible spectroscopy. To this aim, we use the method of continuous variation developed by Paul Job, leading to what we describe as Job scattering plots [36-41]. Here we provide analytical expressions for the Rayleigh ratio obtained from mixed nanoparticle-vesicle aggregates and particle coated with a single bilayer, respectively. Quantitative comparisons with experimental data are also discussed.

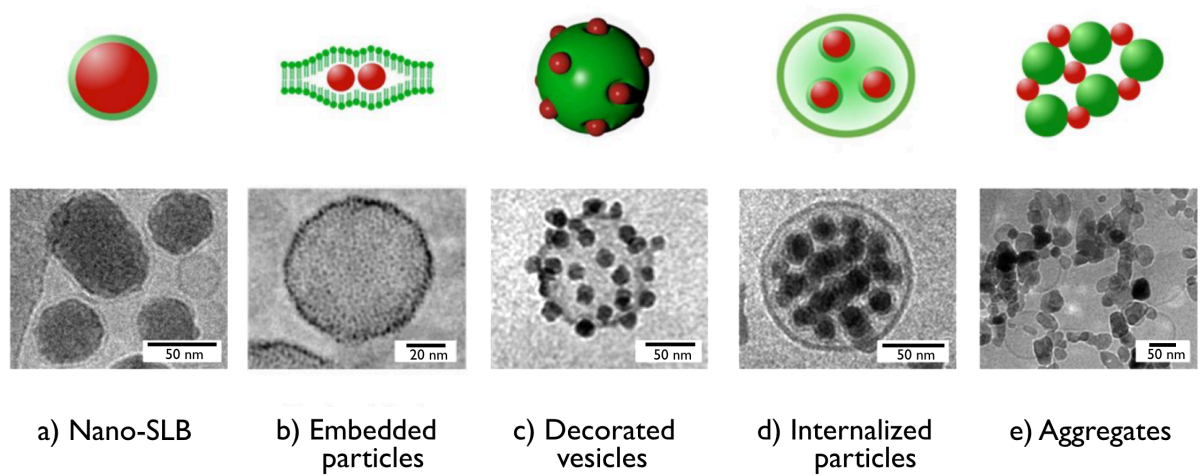

*Figure 1*: *Cryogenic transmission electron microscopy images obtained from nanoparticle-vesicle association. a) Silica nanoparticles coated with a supported lipid bilayer [17]; b) Gold particles embedded within the lipid membrane of a vesicle [31]; c) Silica particles adsorbed at the surface of a vesicles [30]. d) Silica particles internalized inside the lipid compartment [32]. e) Aggregates of ZnO nanoparticles and vesicles [18]. The upper panels provide an illustration for the different structures.*

## II – Experimental

*Nanoparticles*: Aluminum oxide nanoparticles from Disperal® (SASOL, Germany) have the shape of irregular platelets of sizes 40 nm in length and 10 nm in thickness [33]. To obtain homogeneous dispersions, the alumina powder is dissolved in a nitric acid solution (0.4 wt. % in deionized water) at the concentration of 10 g L$^{-1}$ and sonicated for an hour. The particles have an hydrodynamic diameter $D_H$ = 64 nm. The positively charged silica particles were synthetized using the Stöber synthesis. Following the synthesis, the silica were functionalized by amine groups, resulting in a positive coating [17,23,42]. Aminated silica were synthesized at 40 g L$^{-1}$ and diluted with DI-water at pH 5. The hydrodynamic and geometric diameters were determined at $D_H$ = 60 nm and $D_{TEM}$ = 41.2 nm. Negative silica particles (trade name CLX®) were purchased from Sigma Aldrich at the concentration of 450 g L$^{-1}$. The batch was diluted down to 50 g L$^{-1}$ and dialyzed against DI-water at pH 9 for two days. The diameters were measured at $D_H$ = 34 nm and $D_{TEM}$ = 20 nm [23]. The particle surface charge densities were determined using the polyelectrolyte assisted charge titration spectrometry [39], leading densities of +7.3*e*, +0.62*e* and -0.31*e* nm$^{-2}$ respectively. In the following, the particles are abbreviated Alumina (+), Silica (+) and Silica (-).

*Phospholipids*: Dipalmitoylphosphatidylcholine (DPPC), L-α-Phosphatidyl-DL-glycerol sodium salt from egg yolk lecithin (PG, Sigma-Aldrich, MDL number: MFCD00213550) and 2-Oleoyl-1-palmitoyl-sn-glycero-3-phospho-rac-(1-glycerol) (POPG) were purchased from Sigma-Aldrich. Phospholipids DPPC, PG and POPG were dissolved in methanol, at 10, 10 and 20 g L$^{-1}$ respectively and then mixed in proper amounts for a final weight concentration of 80% / 10% / 10% of DPPC / PG / POPG. The solvent was evaporated under low pressure at 60 ˚C for 30 minutes. The lipid film formed on the bottom of the flask was then rehydrated with the addition of Milli-Q water at 60 ˚C and agitated at atmospheric pressure for another 30 minutes. Milli-Q water was added again to finally obtain a solution at 1 g L$^{-1}$. The lipid vesicles are characterized by a hydrodynamic dimeter $D_H$ = 120 nm and a zeta potential $\zeta$ = - 32 mV. Nanoparticle-vesicle interactions were investigated using a mixing protocol known as the continuous variation method [36,39,41]. Surfactant and particle batches were prepared in the same conditions of pH and concentration and the stock solutions were mixed at different volumetric ratios $X = c_{Ves}/c_{NP}$, where $c_{Ves}$ and $c_{NP}$ are the vesicle and nanoparticle concentrations. For Alumina (+), the pH of the stock dispersions was adjusted at pH 5 to ensure that particles do not aggregate as a result of the pH changes. Silica (+) and Silica (-) were studied at physiological pH.

***Static and Dynamic Light Scattering***: The scattered intensity $I_S$ and the hydrodynamic diameter $D_H$ were obtained from the NanoZS Zetasizer spectrometer (Malvern Instruments). Analytical expressions in the following will be given in terms of the Rayleigh ratio $\mathcal{R}$, which is basically proportional to $I_S$ in the case considered here. The second-order autocorrelation function was analyzed using the cumulant and CONTIN algorithms to determine the average diffusion coefficient $D_C$ of the scatterers. $D_H$ was calculated according to the Stokes-Einstein relation $D_H = k_B T/3\pi\eta D_C$ where $k_B$ is the Boltzmann constant, $T$ the temperature and $\eta$ the solvent viscosity. The hydrodynamic diameters provided here are the second coefficients in the cumulant analysis described as $Z_{Ave}$. Measurements were performed in triplicate at 25 °C and 37 °C after an equilibration time of 120 s, yielding experimental uncertainties better than 10% in both intensity and diameter.

***Electrophoretic mobility and zeta potential***: Laser Doppler velocimetry using the phase analysis light scattering mode and detection at an angle of 16° was used to carry out the electrokinetic measurements of electrophoretic mobility and zeta potential with the Zetasizer Nano ZS equipment (Malvern Instruments, UK). Zeta potential was measured after a 120 s equilibration at 25 °C.

# III – Results and discussion

## III.1 – Job scattering plots

In 1928, Paul Job developed the method of continuous variation to determine the stoichiometry of binding (macro)molecular species in solutions, providing information about the equilibrium complexes. We have adapted this technique to study interactions in soft condensed matter using small-angle scattering techniques. In the cases of coacervation or microphase separation , the Job scattering technique allows to screen large domains of phase diagrams and to detect phase boundaries [43]. In the cases of protein forming corona, of polymer or lipid adsorption on nanoparticles, the method is quantitative and provide some key features of the association, e.g. the stoichiometry, the layer thickness and density [41,44]. In this work, emphasis is put on static light scattering and the modeling of attractive interaction between nanoparticles and lipids. This approach leads to analytical expressions for the scattering intensity during aggregate or SLB formation.

More specifically, we are concerned with ternary phase diagrams for which the total active concentration $c = c_{NP} + c_{Ves}$ is constant and the ratio between the two concentrations is varying continuously according to $X = c_{Ves}/c_{NP}$, where $c_{NP}$ and $c_{Ves}$ are the nanoparticle and vesicle concentrations, respectively. In practice, $c$ is held in the range $0.01 - 10$ g L$^{-1}$ and $X = 10^{-3} - 10^3$. This technique has several advantages, one of them being that the solutions are in the dilute regime and that the Debye-Gans theory applies to all solutions [45]. The approach also relies on the fact that the scattering intensity arising from different species is additive, leading to:

$$\mathcal{R}(q,c,X) = \sum_i K_i c_i(X) \left[ \frac{1}{M_w^i}\left(1 + \frac{q^2 R_{G,i}^2}{3}\right) + 2A_{2,i}c_i(X) \right]^{-1} \quad (1)$$

where the index $i$ refers to the different types of scatterer. In this work, 4 types of scatterers are considered: engineered nanoparticles, lipid vesicles, hybrid aggregates and supported lipid bilayers. In Eq. 1, $K_i$ is the scattering contrast coefficient, $M_w^i$ the weight-averaged molecular weight, $R_{G,i}$ the radius of gyration and $A_{2,i}$ is the second virial coefficient. In the following, the form factor $(1 - q^2 R_G^2/3)$ and the interaction contribution $2A_2c$ will be neglected for sake of simplicity, leading for the Rayleigh ratio an expression of the form:

$$\mathcal{R}(c,X) = \sum_i K_i M_w^i c_i(X) \quad (2)$$

The scattering intensity arising from nanoparticle and vesicle mixed solutions is now provided for three basic behaviors, the case of non-interacting species, the aggregate formation (Fig. 1e) and the nano-SLB (Fig. 1a).

## III.2 – Non-interacting species

For the trivial case where nanoparticles and vesicles do not interact, the scattering intensity is the sum of the $X = 0$ and $X = \infty$ intensities weighted by their actual concentrations $c_{NP}(X) = c/(1 + X)$ and $c_{Ves}(X) = cX/(1 + X)$, leading to:

$$\mathcal{R}_{NI}(c,X) = K_{NP} M_w^{NP} c \frac{1}{1+X} + K_{Ves} M_w^{Ves} c \frac{X}{1+X} \quad (3)$$

Eq. 2 is a slowly varying function of $X$ ranging between the Rayleigh ratio of the nanoparticles ($X = 0$) and that of the vesicles ($X = \infty$). Examples of $\mathcal{R}_{NI}(X)$-behavior are shown in Figs. 2 for 40 nm nanoparticles and 100/200 nm vesicles (continuous lines in grey). More details about these calculations are provided in the next section.

## III.2 – Nanoparticle-vesicle hybrid aggregates

Here we consider that the particle-vesicle interaction is attractive and leads to the formation of mixed aggregates. The model is general and does not specify the interaction type. As the scattering varies linearly with the weight-averaged molecular weight of the scatterers, the presence of aggregates will lead to an excess scattering compared to the non-interacting case (Eq. 3). For sake of simplicity, it is assumed that $(m,n)$-aggregates are formed and composed of $m$ nanoparticles and $n$ vesicles. The aggregate molecular weight thus reads $M_w^{Agg} = mM_w^{NP} + nM_w^{Ves}$. This later equation has an important consequence, namely that at the critical stoichiometric ratio $X_C$, all the particles and vesicles put in the solution will be in aggregates, leading to the relationships:

$$X_C = \frac{n}{m} \frac{M_w^{Ves}}{M_w^{NP}} \qquad (4)$$

$$M_w^{Agg} = nM_w^{Ves} \frac{X_C + 1}{X_C} \qquad (5)$$

The above results also suggest that the overall mixing diagram can be decomposed in two regions: for $X < X_C$ nanoparticles are in excess and all added vesicles are consumed in the aggregate formation and for $X > X_C$, the vesicles are the main component and coexist with hybrid aggregates in solutions. In the first regime, the hybrid aggregates are in equilibrium with free nanoparticles, and in the second with free vesicles. Counting the different species as a function of $X$ leads to the following expressions [38,46]:

**Regime 1 → $X < X_C$**

$$\mathcal{R}_{Agg}(c, X < X_C) = K_{NP} M_w^{NP} c \frac{X_C - X}{X_C(1+X)} + K_{Agg} M_w^{Agg} c \frac{X(1+X_C)}{X_C(1+X)} \qquad (6a)$$

**Regime 2 → $X > X_C$**

$$\mathcal{R}_{Agg}(c, X > X_C) = K_{Agg} M_w^{Agg} c \frac{1+X_C}{1+X} + K_{Ves} M_w^{Ves} c \frac{X - X_C}{1+X} \qquad (6b)$$

The previous equations predict the scattering intensity in the case of aggregate formation. They have been estimated for different sets of parameters, as shown in Fig. 2. Tests were performed using two values of vesicular diameter, 100 and 200 nm and three scattering contrast conditions, $(K_{NP}, K_{Ves}, K_{Agg})$ = (1, 0.5, 0.8), (1, 1, 1) and (1, 2, 1.2). The $K$-values were selected to encompass a broad range of refractive index properties ($K$ is indeed proportional to the square of the refractive index increment $dn_s/dc$) [45]. For these calculations, the stoichiometry was assumed to be 10 particles per vesicle. The results shown in Fig. 2 all display a marked maximum centered on $X_C$ = 0.16 ($D_{Ves}$ = 100 nm) and 0.64 ($D_{Ves}$ = 200 nm). For the two sizes, the scattering peaks increase with the lipid contrast. Also shown in the panels are the non-interacting predictions obtained from Eq. 3. Note that these calculations could be easily extended to other types of assemblies such as particles embedded in the membrane, vesicles decorated with nanoparticles or particles internalized inside the membrane compartment. In these latter cases, the molecular weight $M_w^{Agg}$ and the stoichiometry should be adjusted to take into account the modeled structure.

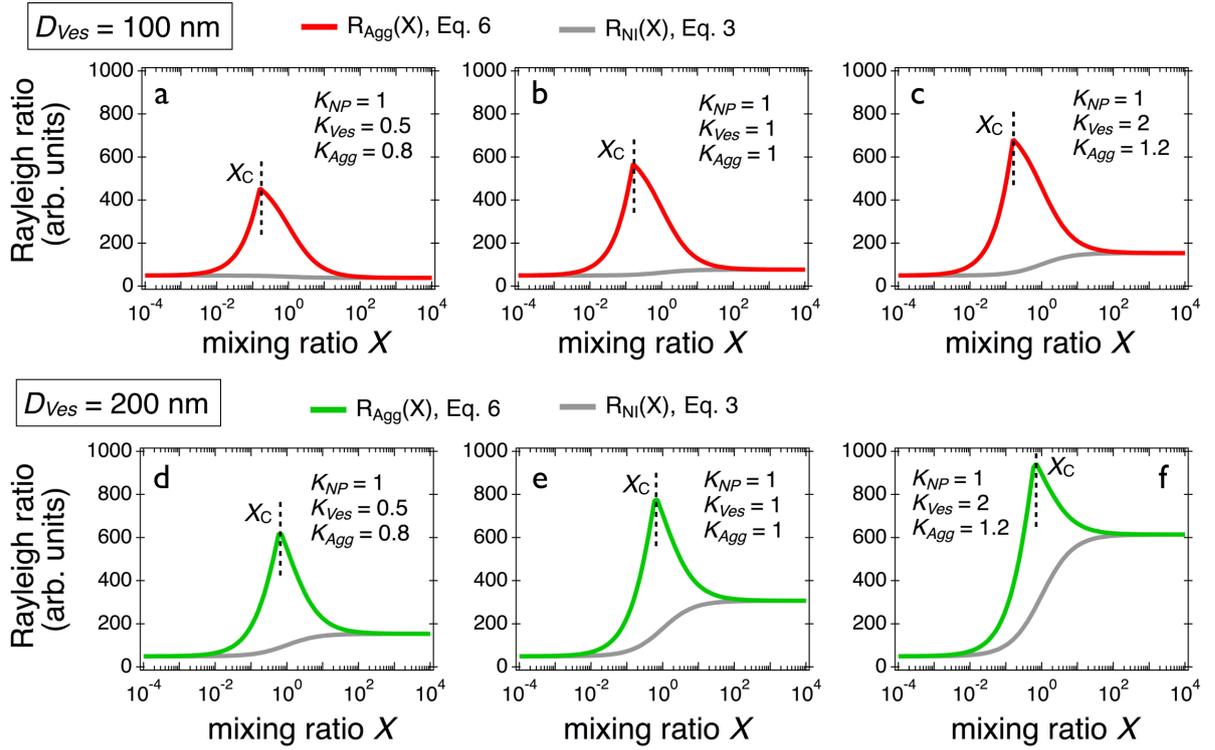

*Figure 2: a), b) and c) Scattering intensity obtained for 40 nm silica nanoparticles and 100 nm lipid vesicles mixed solutions in the case of aggregate formation using three different sets of contrasts, $(K_{NP}, K_{Ves}, K_{Agg}) = (1, 0.5, 0.8), (1, 1, 1)$ and $(1, 2, 1.2)$, respectively. The curves in red refer to $(m = 10, n = 1)$-aggregates (Eqs. 6) whereas those in grey arise from non-interacting species (Eq. 3). $m$ and $n$ denotes the number of particles and of vesicles in the aggregates. d), e) and f) Scattering intensity obtained for 40 nm silica nanoparticles and 200 nm lipid vesicles mixed solutions with the same set of contrasts as above.*

### III.3 – Nano- supported lipid bilayer (nano-SLB)

For nano-SLB, there also exists a critical mixing ratio $X_C$ for which all the particles are covered with a single lipid bilayer and form a supported lipid bilayer (Fig. 1a). At $X_C$, the nanoparticle and vesicle surface area concentrations are equal, so :

$$X_C = \frac{A_{NP}}{A_{Ves}} \qquad (7)$$

where $A_{NP}$ and $A_{Ves}$ denote the specific surface areas for nanoparticles and vesicles respectively. For particles of diameter $D_{NP}$ and mass density $\rho$, one has $A_{NP} = 6/\rho D_{NP}$. For vesicles, the specific surface area reads $A_{Ves} = s_L \mathcal{N}_A / 2M_n^L$, where $s_L$ is the area per polar head, $\mathcal{N}_A$ the Avogadro number and $M_n^L$ the lipid number-average molecular weight. In the following simulation, we discuss the case of 40 nm silica particles (density $\rho$ = 2200 kg m$^{-3}$) and of DPPC lipids of area per polar head $s_L$ = 0.6 nm$^{-2}$ (molecular weight $M_n^L$ = 734 g mol$^{-1}$). With this assumption, one gets $A_{NP}$ = 6.8×10$^5$ cm$^2$ g$^{-1}$ and $A_{Ves}$ = 2.4×10$^6$ cm$^2$ g$^{-1}$, and a value for $X_C$ = 0.28 that does not depend on the vesicle size. The total scattering cross section arises from the sum of the coated and uncoated particle contributions. The scattering intensity then reads [23]:

**Regime 1 → $X < X_C$**

$$\mathcal{R}_{SLB}(c, X < X_c) = K_{NP} \, M_w^{NP} c \frac{(X_C - X)}{X_C(1+X)} + K_{SLB} \, M_w^{SLB} c \frac{X(1+X_C)}{X_C(1+X)} \qquad (8a)$$

**Regime 2 → $X > X_C$**

$$\mathcal{R}_{SLB}(c, X > X_c) = K_{SLB} \, M_w^{SLB} c \frac{1+X_C}{1+X} + K_{Ves} \, M_w^{Ves} c \frac{(X-X_C)}{(1+X)} \qquad (8b)$$

where $M_w^{SLB} = M_w^{NP}(1 + X_C)$ and $K_{SLB}$ the scattering contrast for the SLB-coated particles. Note the similarities of the $X$-dependences in Eqs. 6 and Eqs. 8. Differences however can be found in the determination of $X_C$, in the contrast coefficients and in the molecular weights. The continuous lines in red (resp. in green) in Figs. 3 were obtained from 40 nm particles and 100 (resp. 200 nm) vesicles using the scattering contrast conditions of Fig. 2, ($K_{NP}, K_{Ves}, K_{Agg}$) = (1, 0.5, 0.8), (1, 1, 1) and (1, 2, 1.2). In all the examples considered, the nano-SLB scattering lies below that of the non-interacting species (Eq. 3), which are indicated as continuous lines in grey in each figure. From this, it can be concluded that Job scattering plots display very different features for aggregate and nano-SLB formation. On one side, embedded in the membrane, vesicles decorated with nanoparticles or particles internalized inside the membrane or aggregates give rise to an increase in light scattering and on the other side nano-SLBs are characterized by a decrease of the intensity around a critical ratio. Note that under certain conditions (e.g. in Fig.

3b), this decrease is modest and would not be detectable experimentally. An interaction strength parameter $S_{Int}$ can be defined from the integral beneath the Rayleigh ratio scattering curve $\mathcal{R}(X)$ relative to that of the non-interacting model, such as [23]:

$$S_{Int} = \frac{1}{\mathcal{R}(c, X=0)\, c} \int_0^\infty \big(\mathcal{R}(c,X) - \mathcal{R}_{NI}(c,X)\big) dX \qquad (9)$$

To allow comparison between different particulate systems, the integral is normalized with respect to the nanoparticle intensity and to concentration. According to Eq. 9, the aggregate and SLB formations are characterized by $S_{Int} > 0$ and $S_{Int} < 0$ respectively, whereas the non-interacting systems give $S_{Int} = 0$, a result that should also help in identifying the nature of the association [23].

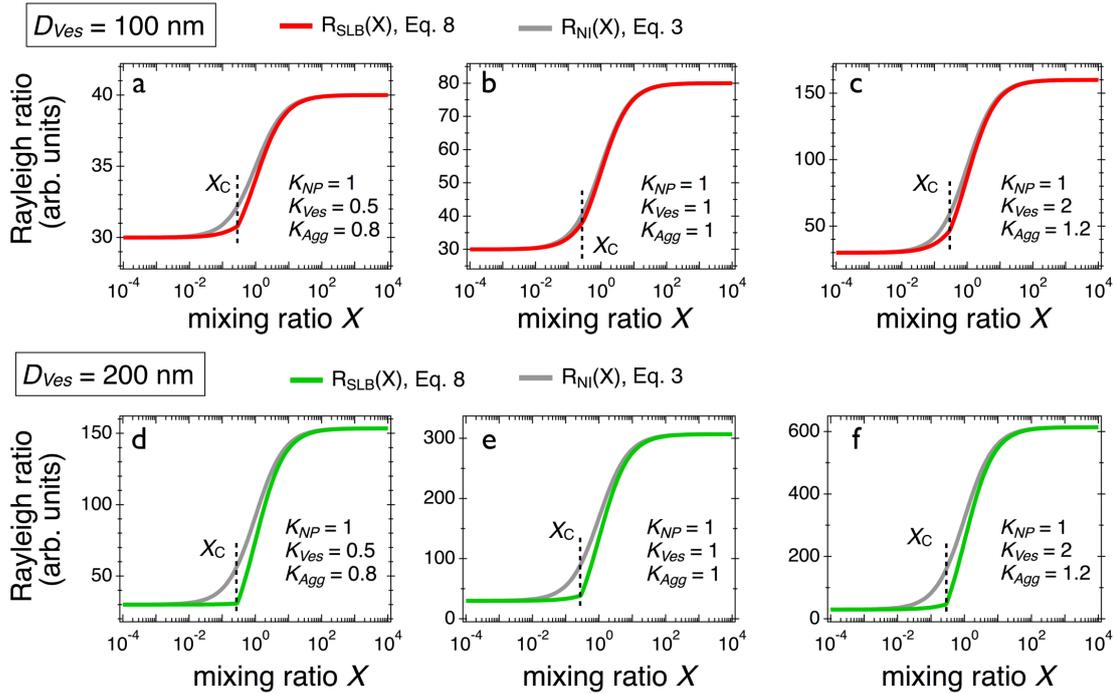

*Figure 3:* a), b) and c) Scattering intensity obtained for 40 nm silica nanoparticles and 100 nm lipid vesicles forming supported lipid bilayers using different contrasts, $(K_{NP}, K_{Ves}, K_{Agg})$ = (1, 0.5, 0.8), (1, 1, 1) and (1, 2, 1.2), respectively. The curves in red refer to nano-SLB (Eqs. 8), whereas the curves in grey arise from non-interacting model (Eq. 3). d), e) and f) Same as before for 200 nm vesicles. Note that for the different conditions, the scattering intensity for the SLB formation remains below that of the non-interacting species.

## III.4 – Comparison with experiments

In this part we study the interactions between nanoparticles and vesicles and compare Job scattering plots obtained experimentally with the above predictions. For the experimental studies, we used vesicles made from a mixture of DPPC, PG and POPG lipids in a ratio 80:10:10 synthesized *via* the evaporation-rehydratation technique. For the particles, we used the aluminum and silicon oxide particles, Alumina (+), Silica (+) and Silica (-) of sizes around 40 nm. The data shown in Figs. 4 reveal that the scattering intensities $\mathcal{R}(X)$ are characterized by strong maxima, indicating the formation of aggregated structures. These peaks correlate well with the hydrodynamic diameters $D_H(X)$ which also pass through a maximum at $X_C$. The scattering intensity for the Alumina (+) was successfully adjusted using Eqs. 6, leading a stoichiometry of 2 particles per vesicle at the two temperatures investigated. For Silica (+) the scattering maxima are less prominent and shifted to lower X ($X_C$ = 0.2), leading to a stoichiometry of 10 particles per vesicle. For Silica (-), the scattering intensity varies monotonously as a function of $X$, in agreement with the non-interacting prediction from Eq. 3. The results from Fig. 4 suggest that the nanoparticle-vesicle aggregate formation is driven by electrostatic interaction. For same charge systems, aggregation is not observed. Finally, none of the nanoparticles tested here display light scattering signatures characteristic of SLBs.

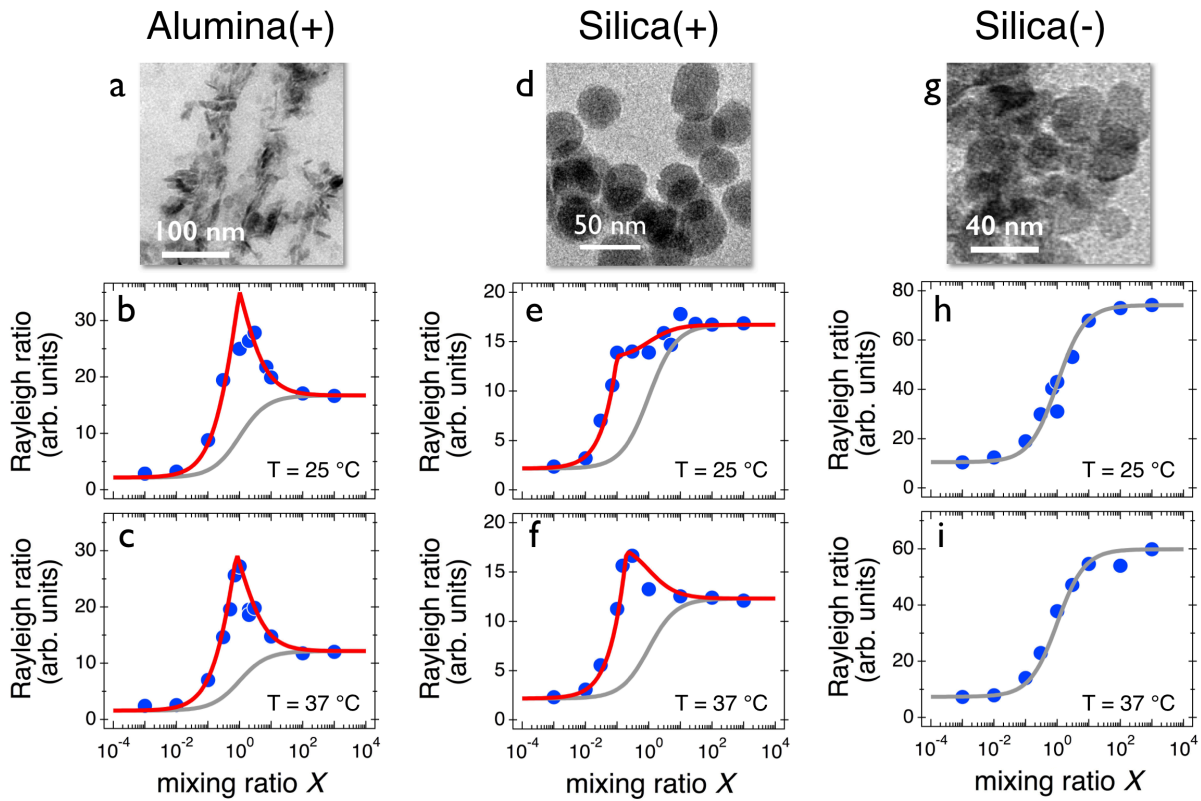

*Figure 4*: **a, d, e)** Transmission electron microscopy images of alumina, positive and negative silica particles used in this study. The particles are dubbed Alumina (+), Silica (+) and Silica (-), respectively. **b and c)** Rayleigh ratio of alumina-vesicle dispersions as a function of the mixing ratio at $T = 25$ °C and $T = 37$ °C respectively. The continuous lines in red are from Eqs. 6, indicating the formation of mixed aggregates and those in grey from Eq. 3. **e and f)** Same as in Fig. 4b and 4c for Silica (+) particles. **h and i)** Rayleigh ratio of negatively charged silica-vesicle dispersions as a function of the mixing ratio at $T = 25$ °C and $T = 37$ °C respectively. The continuous lines in grey are from Eq. 3, indicating no interaction.

# IV – Conclusion

In this work we study the interaction of engineering nanoparticles with lipid vesicles and search for prominent features pertaining to their scattering properties. The first goal is to provide tools to characterize the different types of structures resulting from synthetic/biological membranes and nanomaterials, an issue that is relevant in many biophysical applications. The second objective consists in writing down quantitative predictions for the scattering of dilute solutions, allowing to differentiate between the formation of aggregates and that of supported lipid bilayer. The expressions for the scattering cross-sections are simple and analytical, and they show the relevance of the Job scattering plot approach, as different association scenarios can be discriminated. It is found for instance that the SLB formation is associated with a decrease of the scattering intensity, whereas the aggregate formation is associated with an increase in light scattering. The models proposed are also implementable as the form factor, the dispersity or the interaction of the particles and vesicles can be taken into account into the equations. The result that emerges from the experiments on alumina and silica particles is that electrostatics plays an important role in the association, but is not sufficient to induce the formation of supported lipid bilayers.

# Acknowledgments

ANR (Agence Nationale de la Recherche) and CGI (Commissariat à l'Investissement d'Avenir) are gratefully acknowledged for their financial support of this work through Labex SEAM (Science and Engineering for Advanced Materials and devices) ANR 11 LABX 086, ANR 11 IDEX 05 02. This research was supported in part by the Agence Nationale de la Recherche under the contracts: ANR-13-BS08-0015 (PANORAMA), ANR-12-CHEX-0011 (PULMONANO) and ANR-15-CE18-0024-01 (ICONS) and by Solvay.

# References

1.      Brook, R. D.; Rajagopalan, S.; Pope, C. A.; Brook, J. R.; Bhatnagar, A.; Diez-Roux, A. V.; Holguin, F.; Hong, Y. L.; Luepker, R. V.; Mittleman, M. A.; Peters, A.; Siscovick, D.; Smith, S. C.; Whitsel, L.; Kaufman, J. D. Particulate matter air pollution and cardiovascular disease: An update to the scientific statement from the American Heart Association. *Circulation* **2010,** *121* (21), 2331-2378. doi:10.1161/CIR.0b013e3181dbece1
2. Xia, T.; Zhu, Y. F.; Mu, L. N.; Zhang, Z. F.; Liu, S. J. Pulmonary diseases induced by ambient ultrafine and engineered nanoparticles in twenty-first century. *Nation. Sci. Rev.* **2016,** *3* (4), 416-429. doi:10.1093/nsr/nww064
3. Bajaj, P.; Harris, J. F.; Huang, J. H.; Nath, P.; Iyer, R. Advances and Challenges in Recapitulating Human Pulmonary Systems: At the Cusp of Biology and Materials. *ACS Biomater. Sci. Eng* **2016,** *2* (4), 473-488. doi:10.1021/acsbiomaterials.5b00480
4. Puisney, C.; Baeza-Squiban, A.; Boland, S. Mechanisms of Uptake and Translocation of Nanomaterials in the Lung. In *Cellular and Molecular Toxicology of Nanoparticles*, Saquib, Q.; Faisal, M.; Al-Khedhairy, A. A.; Alatar, A. A., Eds.; Springer International Publishing: Cham, 2018, pp 21-36.
5. Lopez-Rodriguez, E.; Perez-Gil, J. Structure-function relationships in pulmonary surfactant membranes: From biophysics to therapy. *Biochimica Et Biophysica Acta-Biomembranes* **2014,** *1838* (6), 1568-1585. doi:10.1016/j.bbamem.2014.01.028
6. Wustneck, R.; Perez-Gil, J.; Wustneck, N.; Cruz, A.; Fainerman, V. B.; Pison, U. Interfacial properties of pulmonary surfactant layers. *Adv. Colloids Interface Sci.* **2005,** *117* (1-3), 33-58. doi:10.1016/j.cis.2005.05.001
7. Rascol, E.; Devoisselle, J. M.; Chopineau, J. The relevance of membrane models to understand nanoparticles-cell membrane interactions. *Nanoscale* **2016,** *8* (9), 4780-4798. doi:10.1039/c5nr07954c
8. Michel, R.; Gradzielski, M. Experimental Aspects of Colloidal Interactions in Mixed Systems of Liposome and Inorganic Nanoparticle and Their Applications. *Int. J. Mol. Sci.* **2012,** *13* (9), 11610-11642. doi:10.3390/ijms130911610
9. Froehlich, E. The role of surface charge in cellular uptake and cytotoxicity of medical nanoparticles. *Int. J. Nanomed.* **2012,** *7*, 5577-5591. doi:10.2147/ijn.s36111
10. Troutier, A.-L.; Ladaviere, C. An overview of lipid membrane supported by colloidal particles. *Adv. Colloids Interface Sci.* **2007,** *133* (1), 1-21. doi:10.1016/j.cis.2007.02.003
11. Zhang, S. L.; Gao, H. J.; Bao, G. Physical Principles of Nanoparticle Cellular Endocytosis. *ACS Nano* **2015,** *9* (9), 8655-8671. doi:10.1021/acsnano.5b03184
12. Conner, S. D.; Schmid, S. L. Regulated portals of entry into the cell. *Nature* **2003,** *422* (6927), 37-44. doi:10.1038/nature01451
13. Bahrami, A. H.; Raatz, M.; Agudo-Canalejo, J.; Michel, R.; Curtis, E. M.; Hall, C. K.; Gradzielski, M.; Lipowsky, R.; Weikl, T. R. Wrapping of nanoparticles by membranes. *Adv. Colloids Interface Sci.* **2014,** *208* (0), 214-224. doi:10.1016/j.cis.2014.02.012
14. Dasgupta, S.; Auth, T.; Gompper, G. Shape and Orientation Matter for the Cellular Uptake of Nonspherical Particles. *Nano Letters* **2014,** *14* (2), 687-693. doi:10.1021/nl403949h
15. Deserno, M.; Gelbart, W. M. Adhesion and wrapping in colloid-vesicle complexes. *J. Phys. Chem. B* **2002,** *106* (21), 5543-5552. doi:10.1021/jp0138476
16. Mornet, S.; Lambert, O.; Duguet, E.; Brisson, A. The formation of supported lipid bilayers on silica nanoparticles revealed by cryoelectron microscopy. *Nano Letters* **2005,** *5* (2), 281-285. doi:10.1021/nl048153y
17. Mousseau, F.; Puisney, C.; Mornet, S.; Le Borgne, R.; Vacher, A.; Airiau, M.; Baeza-Squiban, A.; Berret, J. F. Supported pulmonary surfactant bilayers on silica nanoparticles: formulation, stability and impact on lung epithelial cells. *Nanoscale* **2017,** *9* (39), 14967-14978. doi:10.1039/c7nr04574c
18. Liu, J. W. Interfacing Zwitterionic Liposomes with Inorganic Nanomaterials: Surface Forces, Membrane Integrity, and Applications. *Langmuir* **2016,** *32* (18), 4393-4404. doi:10.1021/acs.langmuir.6b00493

19. Sachan, A. K.; Harishchandra, R. K.; Bantz, C.; Maskos, M.; Reichelt, R.; Galla, H. J. High-Resolution Investigation of Nanoparticle Interaction with a Model Pulmonary Surfactant Monolayer. *ACS Nano* **2012,** *6* (2), 1677-1687. doi:10.1021/nn204657n

20. Pera, H.; Nolte, T. M.; Leermakers, F. A. M.; Kleijn, J. M. Coverage and Disruption of Phospholipid Membranes by Oxide Nanoparticles. *Langmuir* **2014,** *30* (48), 14581-14590. doi:10.1021/la503413w

21. Savarala, S.; Ahmed, S.; Ilies, M. A.; Wunder, S. L. Formation and Colloidal Stability of DMPC Supported Lipid Bilayers on SiO2 Nanobeads. *Langmuir* **2010,** *26* (14), 12081-12088. doi:10.1021/la101304v

22. Le Bihan, O.; Bonnafous, P.; Marak, L.; Bickel, T.; Trepout, S.; Mornet, S.; De Haas, F.; Talbot, H.; Taveau, J. C.; Lambert, O. Cryo-electron tomography of nanoparticle transmigration into liposome. *J. Struct. Biol.* **2009,** *168* (3), 419-425. doi:10.1016/j.jsb.2009.07.006

23. Mousseau, F.; Berret, J. F. The role of surface charge in the interaction of nanoparticles with model pulmonary surfactants. *Soft Matter* **2018,** *14* (28), 5764-5774. doi:10.1039/c8sm00925b

24. Waisman, D.; Danino, D.; Weintraub, Z.; Schmidt, J.; Talmon, Y. Nanostructure of the aqueous form of lung surfactant of different species visualized by cryo-transmission electron microscopy. *Clin. Physiol. Funct. Imaging* **2007,** *27* (6), 375-380. doi:10.1111/j.1475-097X.2007.00763.x

25. Schleh, C.; Muhlfeld, C.; Pulskamp, K.; Schmiedl, A.; Nassimi, M.; Lauenstein, H. D.; Braun, A.; Krug, N.; Erpenbeck, V. J.; Hohlfeld, J. M. The effect of titanium dioxide nanoparticles on pulmonary surfactant function and ultrastructure. *Respir. Res.* **2009,** *10*, 90. doi:10.1186/1465-9921-10-90

26. Bernardino de la Serna, J.; Vargas, R.; Picardi, V.; Cruz, A.; Arranz, R.; Valpuesta, J. M.; Mateu, L.; Perez-Gil, J. Segregated ordered lipid phases and protein-promoted membrane cohesivity are required for pulmonary surfactant films to stabilize and protect the respiratory surface. *Faraday Discuss.* **2013,** *161* (0), 535-548. doi:10.1039/C2FD20096A

27. Curstedt, T.; Halliday, H. L.; Speer, C. P. A Unique Story in Neonatal Research: The Development of a Porcine Surfactant. *Neonatology* **2015,** *107* (4), 321-329. doi:10.1159/000381117

28. De Backer, L.; Braeckmans, K.; Stuart, M. C. A.; Demeester, J.; De Smedt, S. C.; Raemdonck, K. Bio-inspired pulmonary surfactant-modified nanogels: A promising siRNA delivery system. *J. Control. Release* **2015,** *206*, 177-186. doi:10.1016/j.jconrel.2015.03.015

29. Wohlleben, W.; Driessen, M. D.; Raesch, S.; Schaefer, U. F.; Schulze, C.; von Vacano, B.; Vennemann, A.; Wiemann, M.; Ruge, C. A.; Platsch, H.; Mues, S.; Ossig, R.; Tomm, J. M.; Schnekenburger, J.; Kuhlbusch, T. A. J.; Luch, A.; Lehr, C. M.; Haase, A. Influence of agglomeration and specific lung lining lipid/protein interaction on short-term inhalation toxicity. *Nanotoxicology* **2016,** *10* (7), 970-980. doi:10.3109/17435390.2016.1155671

30. Michel, R.; Plostica, T.; Abezgauz, L.; Danino, D.; Gradzielski, M. Control of the stability and structure of liposomes by means of nanoparticles. *Soft Matter* **2013,** *9* (16), 4167-4177. doi:10.1039/c3sm27875a

31. Rasch, M. R.; Rossinyol, E.; Hueso, J. L.; Goodfellow, B. W.; Arbiol, J.; Korgel, B. A. Hydrophobic Gold Nanoparticle Self-Assembly with Phosphatidylcholine Lipid: Membrane-Loaded and Janus Vesicles. *Nano Letters* **2010,** *10* (9), 3733-3739. doi:10.1021/nl102387n

32. Michel, R.; Kesselman, E.; Plostica, T.; Danino, D.; Gradzielski, M. Internalization of Silica Nanoparticles into Fluid Liposomes: Formation of Interesting Hybrid Colloids. *Angew. Chem.-Int. Edit.* **2014,** *53* (46), 12441-12445. doi:10.1002/anie.201406927

33. Mousseau, F.; Le Borgne, R.; Seyrek, E.; Berret, J.-F. Biophysicochemical Interaction of a Clinical Pulmonary Surfactant with Nanoalumina. *Langmuir* **2015,** *31* (26), 7346-7354. doi:10.1021/acs.langmuir.5b01639

34. Reviakine, I.; Johannsmann, D.; Richter, R. P. Hearing what you cannot see and visualizing what you hear: Interpreting quartz crystal microbalance data from solvated interfaces. *Anal. Chem.* **2011,** *83* (23), 8838-8848. doi:10.1021/ac201778h

35. Dubochet, J. On the Development of Electron Cryo-Microscopy (Nobel Lecture). *Angewandte Chemie (International ed. in English)* **2018**, doi:10.1002/anie.201804280

36. Job, P. Studies on the formation of complex minerals in solution and on their stability. *Ann. Chim. France* **1928,** *9*, 113-203.

37. Renny, J. S.; Tomasevich, L. L.; Tallmadge, E. H.; Collum, D. B. Method of Continuous Variations: Applications of Job Plots to the Study of Molecular Associations in Organometallic Chemistry. *Angew. Chem.-Int. Edit.* **2013,** *52* (46), 11998-12013. doi:10.1002/anie.201304157

38. Fresnais, J.; Lavelle, C.; Berret, J.-F. Nanoparticle Aggregation Controlled by Desalting Kinetics. *J. Phys. Chem. C* **2009,** *113* (37), 16371-16379. doi:10.1021/jp904665u

39. Mousseau, F.; Vitorazi, L.; Herrmann, L.; Mornet, S.; Berret, J. F. Polyelectrolyte assisted charge titration spectrometry: Applications to latex and oxide nanoparticles. *J. Colloid Interface Sci.* **2016,** *475*, 36-45. doi:http://dx.doi.org/10.1016/j.jcis.2016.04.036

40. Oikonomou, E. K.; Mousseau, F.; Christov, N.; Cristobal, G.; Vacher, A.; Airiau, M.; Bourgaux, C.; Heux, L.; Berret, J. F. Fabric Softener–Cellulose Nanocrystal Interaction: A Model for Assessing Surfactant Deposition on Cotton. *J. Phys. Chem. B* **2017,** *121* (10), 2299-2307. doi:10.1021/acs.jpcb.7b00191

41. Torrisi, V.; Graillot, A.; Vitorazi, L.; Crouzet, Q.; Marletta, G.; Loubat, C.; Berret, J.-F. Preventing Corona Effects: Multiphosphonic Acid Poly(ethylene glycol) Copolymers for Stable Stealth Iron Oxide Nanoparticles. *Biomacromolecules* **2014,** *15* (8), 3171-3179. doi:10.1021/bm500832q

42. Reinhardt, N.; Adumeau, L.; Lambert, O.; Ravaine, S.; Mornet, S. Quaternary Ammonium Groups Exposed at the Surface of Silica Nanoparticles Suitable for DNA Complexation in the Presence of Cationic Lipids. *J. Phys. Chem. B* **2015,** *119* (21), 6401-6411. doi:10.1021/acs.jpcb.5b01834

43. Vitorazi, L.; Ould-Moussa, N.; Sekar, S.; Fresnais, J.; Loh, W.; Chapel, J. P.; Berret, J.-F. Evidence of a two-step process and pathway dependency in the thermodynamics of poly(diallyldimethylammonium chloride)/poly(sodium acrylate) complexation. *Soft Matter* **2014,** *10* (47), 9496-9505. doi:10.1039/c4sm01461h

44. Giamblanco, N.; Marletta, G.; Graillot, A.; Bia, N.; Loubat, C.; Berret, J.-F. Serum Protein-Resistant Behavior of Multisite-Bound Poly(ethylene glycol) Chains on Iron Oxide Surfaces. *ACS Omega* **2017,** *2* (4), 1309-1320. doi:10.1021/acsomega.7b00007

45. Lindner, P.; Zemb, T. *Neutrons, X-rays and Light : Scattering Methods Applied to Soft Condensed Matter*. Elsevier: Amsterdam, 2002.

46. Berret, J.-F. Stoichiometry of electrostatic complexes determined by light scattering. *Macromolecules* **2007,** *40* (12), 4260-4266. doi:10.1021/ma062887a